\documentstyle[11pt,aas2pp4, psfig, tighten]{article}

\def\sp{\hspace{1.5pt}}
\def\etal{{\it et~al.}}
\def\amin{\ifmmode^{\prime}\else$^{\prime}$\fi}
\def\asec{\ifmmode^{\prime\prime}\else$^{\prime\prime}$\fi}

\def\simgt{\lower.5ex\hbox{$\; \buildrel > \over \sim \;$}}
\def\simlt{\lower.5ex\hbox{$\; \buildrel < \over \sim \;$}}

\newcommand\xte{{\it RXTE}}

\newcommand\asca{{\it ASCA}}

\def\sp{\hskip 1.5pt}
\def\snr{\hbox{N157B}}
\def\psr{\hbox{PSR J0537$-$6910}}
\def\lmcpsr{\hbox{PSR B0540$-$69}}
  
\slugcomment{Draft of March 3 1998}

\lefthead{Marshall, Gotthelf, Zhang, Middleditch, \& Wang}
\righthead{The X-ray Pulsar in \snr}

\begin{document}

\title{\large Discovery of an Ultra-fast X-ray Pulsar in the Supernova 
Remnant N157B}

\author{F. E. Marshall\footnote{Laboratory for High Energy
Astrophysics, NASA/GSFC, Greenbelt, MD 20771;
frank.marshall@gsfc.nasa.gov, gotthelf@gsfc.nasa.gov,
zhang@xancus10.gsfc.nasa.gov}, E. V. Gotthelf$^{1,}$\footnote{Universities Space Research Association}, W. Zhang$^1$,
J. Middleditch\footnote{Los Alamos National Laboratory, MS B265,
CIC-19, Los Alamos, NM 87545; jon@lanl.gov} \&
Q. D. Wang\footnote{Dearborn Observatory, Northwestern University,
2131 Sheridan Road, Evanston, IL 60208; wqd@nwu.edu}}

\begin{abstract}
	We present the serendipitous discovery of 16 ms pulsed X-ray
emission from the Crab-like supernova remnant \snr\ in the Large
Magellanic Cloud. This is the fastest spinning pulsar associated with
a supernova remnant (SNR). Observations with the {\it Rossi} X-ray
Timing Explorer (\xte), centered on the field containing SN1987A,
reveal an X-ray pulsar with a narrow pulse profile.  Archival \asca\
X-ray data confirm this detection and locate the pulsar within
$1^{\prime}$ of the supernova remnant \snr, $14^{\prime}$ from
SN1987A. The pulsar manifests evidence for glitch(es) between the
\xte\ and \asca\ observations which span 3.5 years; the mean linear
spin-down rate is ${\dot P} = 5.126 \times 10^{-14} \ {\rm s \
s^{-1}}$. The background subtracted pulsed emission is similar to
other Crab-like pulsars with a power law of photon index of $\sim
1.6$. The characteristic spin-down age ($\sim 5000$ years) is
consistent with the previous age estimate of the SNR.  The inferred
$B$-field for a rotationally powered pulsar is $\sim 1 \times 10^{12}$
Gauss.  Our result confirms the Crab-like nature of \snr ; the pulsar
is likely associated with a compact X-ray source revealed by ROSAT HRI
observations.

\end{abstract}


\section{Introduction}

Crab-like supernova remnants (SNRs) play a critical role in our
understanding of young pulsars and their pulsar wind nebulae.
These rare SNRs which contain central pulsars are distinguished by
their centrally-filled morphologies and non-thermal X-ray spectra.  The
X-ray emission of such SNRs comes predominantly from a synchrotron
nebula powered by an embedded young pulsar (e.g., Seward 1989). Only
three confirmed members of this class were previously known: the Crab
Nebula with its famous 33 ms pulsar PSR B0531$+$21, the SNR B0540$-$693 in
the Large Magellanic Cloud (LMC) with its 50 ms pulsar \lmcpsr,
and the nebula around the 150 ms pulsar PSR B1509$-$58.  Several candidate
Crab-like SNR have been reported. These have similar morphologies and
spectral characteristics but so far lack a detected pulsar.  Young
Crab-like SNRs are expected to evolve into composite SNRs, in which
the thermal emission from shock-heated gas will rival or exceed the
diminishing X-ray radiation from the pulsar-powered synchrotron nebulae.
This evolutionary sequence is not yet firmly established, though,
mainly because of the limited number of samples.

	We report in this Letter the discovery of an ultra fast pulsar
in \snr\ (also known as NGC\sp 2060, SNR 0538$-$69.1, \& 30\sp Dor\sp
B; Henize 1956), confirming the Crab-like nature of this remnant (Wang
\& Gotthelf 1998 and ref. therein). The pulsar was first detected
serendipitously with \xte\ data while searching for a pulsed signal
from nearby SN1987A (Marshall \etal\ 1998). We used archival \asca\
data of the region to confirm the detection of the pulsed emission, to
measure its spin-down rate, and to locate the emission to within the
boundary of \snr.  Based on the \asca\ position, we will refer to the
pulsar as \psr.  In the following, we present our discovery of \psr\
and discuss its properties in the context of other young pulsars.

\section{Observations}

The LMC field containing SN1987A has been frequently sampled by many
X-ray observatories including \xte\ and \asca. There were two
observing runs in 1996 with \xte\ to search for pulsations from
SN1987A (see Observation Log, Table 1).  The first was an
uninterrupted observation of 21 ks on 12 Oct. 1996.  The other run,
which began on 20 Dec. 1996, obtained a total of about 78 ks of good
observing time during an elapsed time of 184 ks.

The \xte\ Observatory (Bradt \etal\ 1993) consists of five co-aligned
collimated detectors known collectively as the Proportional Counter
Array (PCA), a set of crystal scintillation detectors known as the
High Energy X-ray Timing Experiment (HEXTE), and the All-Sky Monitor.
Here we report exclusively on data obtained with the PCA because of
its large effective area ($\sim 6500$ cm$^{2}$ at $10$ keV). The PCA
(Jahoda \etal\ 1996) has collimators that produce a roughly circular
aperture with a $\sim 1^{\circ}$ FWHM response.  There are several
known X-ray sources within the field-of-view (FOV) with flux above $2$
keV, the brightest of which are the black-hole candidate LMC X-1, the
50 ms pulsar \lmcpsr, and \snr, the object of this study.  The PCA
observations used the Good Xenon data mode, which time-tags each
photon with $0.9 \mu$s timing resolution.  For the current analysis,
the absolute timing uncertainty is $\sim 100 \mu$s (Rots \etal\ 1998).
Moderate spectral information is available in the $2-60$ keV energy
band with a resolution of $\sim 16\%$ at 6 keV.  

The \asca\ Observatory (Tanaka \etal\ 1994) acquired three long $\sim
20$ ks observations of the region around SN1987A with sufficient
timing resolution to detect ms pulsations (see Table 1). These were
all obtained in 1993 during the early Performance and Verification phase
of the \asca\ mission. Two of the observations containing \snr\ were
optimal, placing the SNR close to the optical axis, while the third
data set observed the SNR with reduced efficiency, at $18^{\prime}$
off-axis where the source flux is strongly vignetted.

\asca\ data were acquired with all four co-aligned focal plane
instruments. We used data exclusively from the two Gas scintillation
Imaging Spectrometers (GIS), and only data for which time-tagged
photons were available in the highest time resolution mode ($61 \
\mu\rm{s}$ or $0.488 \ {\rm ms}$ depending on telemetry rate). \asca\
GIS timing measurements are found to have an absolute accuracy of $200
 \ \mu\rm{s}$ in this mode (see Saito \etal\ 1997).  Data from the Solid
State Imaging Spectrometers (SIS) do not have adequate time resolution
for the current study.  The GIS offers $\sim 8\%$ spectral resolution
at 6 keV over its $\sim 1-12$ keV energy band-pass.  Each GIS sits at
the focus of a conical foil mirror and the combination results in a
spatial resolution of $1-3$ arcmin (depending on energy) over the
GIS's $\sim 50^{\prime}$ arcmin diameter active FOV. Point sources can
be located with $\sim 0.6^{\prime}$ accuracy (see Gotthelf 1994).

Both the \xte\ and \asca\ data were edited to exclude times of high
background contamination using the standard processing screening
criteria (``REV2'' for \asca).  This rejects time intervals of South
Atlantic Anomaly passages, Earth occultations, bright Earth limb in
the field-of-view (\asca\ only), and other periods of high particle
activity. For each observation, event data from all detectors were
co-added and the arrival times of each photon corrected to the solar
system barycenter using the JPL DE200 ephemeris.

\section{Timing Results}

For this Letter we have concentrated on \xte\ data collected in the
$2-10$ keV energy band in layer 1 of the PCA, which provides
the best sensitivity for weak sources with a Crab-like spectrum, and
the $\sim 10-25$ keV hard energy band using all three PCA
layers. We also examined data above 30 keV, which are expected to be
dominated by detector background, to verify that there are no similar
periodicities produced by the instrument itself.

We searched for significant signals using $2^{27}$ point Fast Fourier
transforms on data binned in 0.16 ms steps. The 12 Oct 1996
observation produced a $7.5 \sigma$ signal at $\sim 62.055$ Hz in the
hard energy band.  Similar power was found for the $2^{nd}$ and
3$^{rd}$ harmonics. The probability of false detection is $< 1$ in
$10^{10}$. This result is reproduced in the $2-10$ keV band data, 
with a $10 \sigma$ detection in the first three harmonics of
the 62 Hz signal. No other significant power was found with the
exception of frequencies corresponding to the 50 ms pulsar \lmcpsr.  The
new frequencies are not simple harmonics or aliases of the 50 ms
pulsar signal. As we show below, any association with the 50 ms
pulsations can be ruled out conclusively.  We folded the data into 20
bins on periods near 16 ms and computed $\chi^2$ for a model with a
count rate independent of pulse phase.  The largest value for $\chi^2$
of 157 ($7.8 \sigma $) was found for a period of $P = 16.114717$ ms.
The pulsar was also detected during the second \xte\ observing run,
allowing us to derive an initial period derivative.

We examined the \asca\ data for the \xte\ pulsations. \snr\ was
considered the most likely source of the pulsations because it is the
brightest source in the \xte\ FOV after LMC X-1 and B0540$-$693. It is
a barely resolved, moderately bright \asca\ source in the 30 Doradus
star formation region. \snr\ is clearly imaged in the \asca\ GIS as an
isolated source, $\sim 10^{\prime}$ southwest of the X-ray bright R136
region, and $\sim 15^{\prime}$ northwest of the 50 ms pulsar \lmcpsr,
which produced stray light in the \asca\ FOV containing \snr.

We extracted photons from both GIS instruments using a $4^{\prime}$
radius aperture centered on the peak flux from \snr\ and constructed a
periodogram as with the \xte\ data. For each trial period, we folded
the data into 10 bins to optimize the signal in the photon noise
dominated data and computed the $\chi^2$ of the resultant
profile. From each \asca\ observation we conducted a high resolution
blind search of the data for significant power about a range of $\pm
0.5 \ \rm{ms}$ around the expected period. To search for short duty
cycle pulsations in period-space we oversampled the period resolution
by a factor of 20, using increments of $0.05 \times P^2/T$, were $T$
is the observation duration, and $P$ is the test period.

A highly significant signal ($> 12\sigma$) was found during both
on-axis \asca\ observations. The low signal-to-noise September 1993
observation produced a $5\sigma$ result, whose reduced significance is
consistent with the target being placed $18^{\prime}$ off-axis. This
is the only significant peak found from all data sets, over the 
range searched.  We used the change in the period between the two
on-axis observations to bootstrap the period derivative, which
resulted in a search of greater sensitivity. We also find that the
period detection is greatly improved by excluding photons below
$<2$ keV. This suggests that the pulsar signal is significantly
contaminated by un-pulsed, soft thermal flux from the SNR nebula,
which dominates below $2$ keV. The non-detection of the pulsar by Wang
\& Gotthelf (1998) is explained by the narrowness of the peak, the
duration of the observations (typically 90 ks), and the lack of a
period derivative {\it a priori}.

Since it is not possible to maintain phase between any of the four
observation intervals, estimates of the period derivative are based on
the change in period between observations.  Figure 1 shows the period
evolution and the residuals from the best-fit linear model of $P =
0.016113089 \ {\rm s} \pm 1.7$~ns at MJD 50000 (at Earth) and
$\dot{P}= 5.126 \pm 0.002 \times 10^{-14} \ {\rm s \ s^{-1}}$.  Our
estimates of the period uncertainty for each data point in Figure 1
are derived by Monte-Carlo simulation and Maximum Likelihood analysis.
Clearly, there are significant deviations from a constant period
derivative; these are not removed by including a period second
derivative to the fit. This inability to find a single solution for
both sets of observations suggest that one or more glitches in the
period has occurred (see discussion in $\S 4$). The measured period
derivatives during the two epochs are $\dot{P}_{\asca} = 5.154 \pm 0.002
\times 10^{-14} \ {\rm s \ s^{-1}}$ and $\dot{P}_{\xte} = 5.174 \pm 0.002 
\times 10^{-14} \ {\rm s \ s^{-1}}$.

Using the above ephemeris, we generated pulse profiles for the 1993
\asca\ and the 1996 \xte\ combined data sets. The resulting profiles
are displayed in Figure 2. They are well characterized by a single,
narrow peaked pulse with an approximately Gaussian shape of FWHM of
$\sim 1.7$ ms (10\% duty cycle) ms. The profiles appear unchanged
between the 1993 and 1996 observations. The pulsed emission comprises
$9.6\%$ of the total \asca\ counts in the $4^\prime$ radius aperture above
$2$ keV.

To demonstrate that these pulsations are uniquely associated with
\snr, an image of the pulsed emission was generated by subtracting 
the off-pulsed data from the on-pulsed image. Only a single source of
significant emission remained, the pulsed emission, an unresolved
\asca\ point-like source located at the coordinates of the SNR
\snr. We thus reproduce and verify the \xte\ measurement, and
unambiguously locate the origin of the pulsed emission to within an
arc minute of the region encompassing \snr.

In summary, we consider this a conclusive detection of a new pulsar
in \snr, \psr, whose period of 16 ms makes it the fastest known
pulsar associated with a SNR.

\section{Discussion}

The steady increase in the period suggests that the pulsed X-ray
emission is powered by the spin-down energy loss of the pulsar.  We
estimate the spin-down power as ${\dot E} = (2\pi)^2 I{\dot P}/P^{3} =
4.8\times 10^{38} I_{45} {\rm \ ergs \ s^{-1}}$, where $I_{45}$ is the
moment of inertia of the neutron star in units of $10^{45}{\rm g\,
cm^{2}}$. For a pulsar losing rotational energy via magnetic dipole
radiation, ${\dot E} \sim (B_p^2 R^6 \Omega^4)/(6c^3)$,
and the surface magnetic field strength at the pole is $B_{p} \sim 1.0
\times 10^{12}$ G.  Here, $R\sim 10{\rm\ km}$ is the neutron star
radius, $\Omega=2\pi/P$ is the angular velocity of the rotation, and
$c$ is the speed of light.

The pulsed spectral component can be isolated using phase-resolved
spectroscopy. The source photons were folded at the best measured
period and phase dependent spectra constructed. Using the off-pulse
spectrum as background, we fit the on-pulsed spectrum with a simple
absorbed power law model to the \asca\ \& \xte\ data in the $2-20$ keV
band. The best fit photon index ($\chi^2_{\nu} = 0.96 $ for 30 DoF) is
$1.6 \ (1.3 - 2.0)$, which is consistent with the typical value $\sim
1.7$ for other pulsars (Lyne \& Graham-Smith 1990). The large
uncertainties are due to subtracting two spectra with similar number
of counts.  The unabsorbed $2-10$ keV pulsed flux of \psr\ is $\sim
6.7 \pm 0.6 \times 10^{-13} \ {\rm ergs \ s^{-1} \ cm^{-2}}$ (see
Table 1).  The corresponding pulsed luminosity is $\sim 1.7 \times
10^{35} \ {\rm ergs \ s^{-1}} $ (into $4 \pi$) assuming a distance of
47 kpc (Gould 1995).

The derived pulsed luminosity suggests that $\sim 4 \times 10^{-4}$ of
the pulsar's spin down energy is emitted as pulsed X-rays in the
$2-10$ keV band. This luminosity is consistent with the empirical
relation, $L_x \approx 7 \times 10^{26} (B_{12}/P^2_s)^{2.7} {\rm~
ergs~s^{-1}}$, which fits well to the magneto-rotation driven X-ray
emission from previously known pulsars (\"Ogelman 1995), to within the
uncertainty in the power index of $2.7 \pm 0.5$. Here, $B_{12}$ is the
magnetic field in units of $10^{12}$ G and $P^2_s$ is the spin period in
units of seconds.

Compared to other young pulsars known to be associated with Crab-like
SNRs, \psr\ is quite unusual.  Its X-ray pulse profile appears to be
the narrowest among the young pulsars. The Crab pulsar is double
peaked and the profiles of both \lmcpsr\ and PSR B1509$-$58 are very
broad (duty cycle $> 0.3$). The pulse width of \psr, $1.7$ ms,
indicates that the size of the emission region is likely smaller than
about half of the light cylinder radius. A detailed study of the pulse
profile may help to place constraints on pulsar emission models. The
characteristic age of the pulsar is $\tau = P/(2{\dot P}) = 5\times
10^{3}$ yrs.  This age is much greater than any of the other Crab-like
pulsar, yet similar to the age estimate for \snr\ of $\sim 5 \times
10^3$~yrs, based on X-ray measurements of the size and temperature of
the remnant (Wang \& Gotthelf 1998). An upper limit on the age of the
remnant, based on the kinematics of H$\alpha$-emitting gas in the
region (Chu \etal\ 1992), suggests that the remnant may be as old as
$\sim 2 \times 10^4$~yrs.

An accurate determination of the remnant age is critical to estimate
the initial rotation rate of the pulsar. \psr\ rotates twice as fast
as the Crab pulsar, and was likely spinning much more rapidly when it
was born, depending on its assumed braking index and age. Following
Kaspi \etal\ (1997), we plot in Figure 3 the possible initial spin
period vs. age for a range of braking index. If the older age estimate
is correct and if glitches are not important in the overall period
evolution of the pulsar, the braking index would then have to be
unusually small ($\le 1.2$) for an initial period $\ge 1$ ms.  If the
younger value for the age is more likely, the initial spin is then a
few ms, assuming $n \sim 3$, as in the case of the magnetic dipole
model and for other young pulsars. In either case, these values
provide important constraints on neutron star birth spin models.

The phenomena of pulsar glitches is well established and provides
important constrains on the moment of inertia of neutron stars (see
Shapiro \& Teukolsky 1983). These sudden changes in the pulsar spin
period and period derivative are attributed to ``starquakes'', stress
relief between the neutron star crust and the superfluid
core. Monitoring of the Crab and Vela pulsars, as well as other young
pulsars, show that large glitches occur on $\sim $ a few years
timescales.  There is compelling evidence that \psr\ has undergone one
or more glitches between the \asca\ and \xte\ epochs (see figure
1). The integrated magnitude of the glitches, $\Delta \Omega / \Omega
\sim 10^{-6}$, is comparable to that observed from the Vela pulsar,
but substantially larger than any of those observed from the Crab
pulsar. More careful analysis of the data is still needed, however, to
quantify the presence of any glitches and to the distinguish them from
timing noise.

The origin of the un-pulsed X-rays detected from \snr\ in the energy
band $\ge 2$~keV is likely a synchrotron nebula powered by the pulsar
via a relativistic wind (Gallant \& Arons 1994 and references
therein).  Recent work by Wang \& Gotthelf (1998) has shown that \snr\
contains a bright, elongated, and non-thermal X-ray feature whose
origin is still uncertain. Attached to this feature is a compact
source with a spatial extent of $\sim 7^{\prime\prime}$, which they
claim most likely represents the pulsar and its interaction with
surrounding medium. The X-ray spectrum of \snr, which is dominated by
the un-pulsed emission, can be characterized by a power law with an
energy slope $\sim 1.5$ (Wang \& Gotthelf 1998), significantly steeper
than those ($\sim 1.0$) of other Crab-like SNRs (e.g., Asaoka \&
Koyama 1990). The remnant has probably evolved to a stage close to
that of a composite SNR.  This steep spectrum and the elongated
morphology suggest that something unusual may be happening in
\snr. Most likely, relativistic particles are transported from the
pulsar to a radio-emitting wind bubble in a collimated outflow (Wang
\& Gotthelf 1998). A substantial fraction of high energy particles
from the pulsar lose their energy radiatively in a bow shock, rather
than adiabatically through diffusion.

\section{Conclusion}

We have found 16 ms pulsed X-ray emission from the 30 Doradus region
of the LMC.  We have shown that these pulsations are associated
uniquely with the X-ray emission from the SNR \snr.  This remnant is
an important new addition to the class of Crab-like SNRs.  It provides
a rare laboratory for studying both an unusually rapidly-rotating
pulsar and its relativistic wind, as well as the structure and
evolution of a neutron star. Future observations are important to
constrain the age of the pulsar, the period second derivative, and to
measure putative glitches.

\begin{acknowledgements}
{\noindent \bf Acknowledgments} --- We thank the \xte\
and \asca\ teams for making these observations possible. We thank
V. Kaspi for discussion and a careful reading of the manuscript. This
research made use of data obtained through the HEASARC online service,
provided by NASA/GSFC. E.V.G. is supported by USRA under NASA contract
NAS5-32490. Q.D.W. is supported partly by NASA LTSA grant NAG5-6413.
\end{acknowledgements}

\begin{deluxetable}{lccccc}
\small
\tablewidth{0pt}
\tablecaption{Observation Log for \psr\vfill
\label{table I}}
\tablehead{
 \colhead{\hfil Data Set \hfil} & \colhead{Date$^a$}& \colhead{Epoch$^a$} & \colhead{Exposure/Duration$^b$} & 
      \colhead{Period$^b$} & \colhead{Pulsed Flux$^c$} \nl
           & \colhead{(UT)} & \colhead{(MJD)} & \colhead{(ks)} & \colhead{(ms)} & 
         \colhead{\hfil ($\times 10^{-13}$ ergs s$^{-1}$ cm$^{-2}$ ) \hfil} \nl
}

\startdata
\noalign{\vskip 5pt}
 \asca\ & 13 Jun 1993 & 49151.306618 & 30.8/89.4  & 16.1093291 &  $6.4 \pm 0.8$ \nl
 \asca\ & 20 Aug 1993 & 49219.986830 & 32.1/94.8  & 16.1096349 &  $7.1 \pm 0.9$ \nl
 \asca\ & 23 Sep 1993 & 49253.982891 & 35.6/82.8  & 16.1097864 &  $4.1 \pm 2.0$ \nl
 \xte\  & 12 Oct 1996 & 50368.148905 & 21.8/21.8  & 16.1147174 &  $6.8 \pm 1.2$ \nl
 \xte\  & 20 Dec 1996 & 50437.550201 & 78.0/184.5 & 16.1150276 &  $6.7 \pm 0.7$ \nl
\enddata
\tablenotetext{a}{For the start of the observation.}
\tablenotetext{b}{The 90\% confidence level uncertainty on all period measurements is $\sim 0.1$ ns, except for the 20 Dec 1996 measurement, which is 0.04 ns.}
\tablenotetext{c}{The unabsorbed pulsed \xte\ and \asca\ flux using an absorbed power law model of photon index 1.6. The \asca\ GIS flux are derived using photons extracted from $4^{\prime}$ radius aperture centered on the \asca\ \snr\ position. The off-pulse spectrum was used as a background. See text for details.}
\end{deluxetable}

\clearpage

\begin{figure}
\centerline{
\psfig{figure=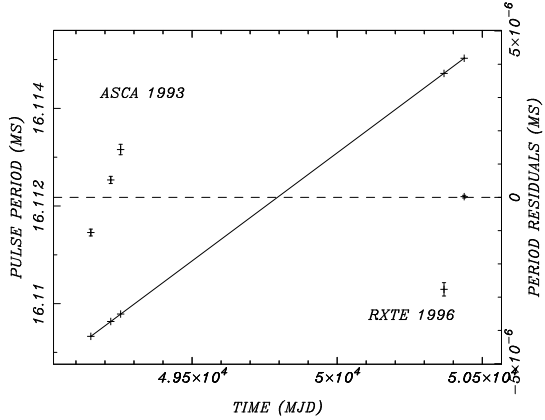,height=2.5truein,angle=270.0,clip=}
}
\caption{The pulse period evolution of \psr\ and its residuals from the best fit model (see text) assuming a linear spin-down trend.}
\end{figure}

\begin{figure}
\centerline{
\psfig{figure=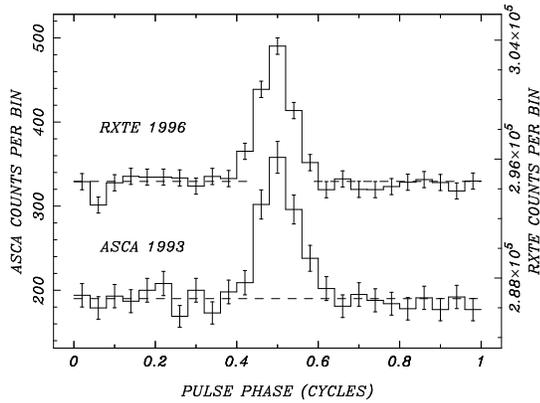,height=2.5truein,angle=270.0,clip=}
}
\caption{The pulse profile of \psr\ in the $2-10$ keV band from the
\xte\ (top) and \asca\ (bottom) observations. The two profiles have
been aligned to place the peak emissions at the 0.5 phase bin. The
relative phases of the two measurements are arbitrary. The \xte\
profile includes 100 ks of data; the \asca\ profile is restricted to
the 37 ks of 64 $\mu \rm{s}$ resolution data.
}
\end{figure}

\begin{figure}
\centerline{
\psfig{figure=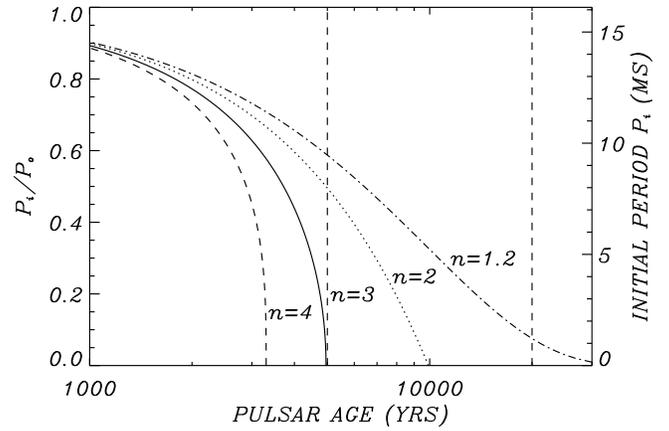,height=2.5truein,angle=0.0}
}
\caption{Predicted initial period of \psr\ as a function of the age of
the pulsar.  Here we have assumed a power-law deceleration model $\dot
P \propto P^{2-n}$ (e.g., Shapiro \& Teukolsky 1983). The four curves
corresponds to different values of the index: 1.2 (dot-dashed), 2
(dotted), 3 (solid), and 4 (dashed).  The vertical lines indicate a
pulsar age of $5 \times 10^{3}$ \& $20 \times 10^{3}$ yrs old.  }
\end{figure}

\end{document}